\documentclass[12pt]{iopart}
\pdfoutput=1

\usepackage{iopams}
\usepackage{bm}
\usepackage{amssymb}
\usepackage{graphicx}
\usepackage{hyperref}
\usepackage{subfigure}

\begin{document}

\title{Belief Propagation and Loop Series on Planar Graphs}

\author{Michael Chertkov$^1$, Vladimir Y. Chernyak$^2$ and Razvan Teodorescu$^1$}
\address{$^1$Theoretical Division and Center for Nonlinear
Studies, Los Alamos National Laboratory, Los Alamos, NM 87545}
\address{$^2$Department of Chemistry, Wayne State University,
5101 Cass Ave,Detroit, MI 48202}
\eads{\mailto{chertkov@lanl.gov}, \mailto{chernyak@chem.wayne.edu},
\mailto{razvan@lanl.gov}}

\begin{abstract}
We discuss a generic model of Bayesian inference with binary
variables defined on edges of a planar graph. The Loop Calculus approach of
\cite{06CCa,06CCb} is used to evaluate the resulting series expansion
for the partition function. We show that, for planar graphs,
truncating the series at single-connected loops reduces, via a map
reminiscent of the Fisher transformation \cite{61Fis}, to evaluating
the partition function of the dimer matching model on an auxiliary planar
graph. Thus, the truncated series can be easily re-summed, using the
Pfaffian formula of Kasteleyn \cite{61Kas}. This allows to identify
a big class of computationally tractable planar models reducible to
a dimer model via the Belief Propagation (gauge) transformation. The
Pfaffian representation can also be extended to the full Loop Series, in which case
the expansion becomes a sum of Pfaffian contributions, each associated with
dimer matchings on an extension to a subgraph of the original graph.
Algorithmic consequences of the Pfaffian representation,
as well as relations to quantum and non-planar models, are
discussed.
\end{abstract}

\submitto{Journal of Statistical Mechanics}

\maketitle

\section{Introduction}

{\it Bayesian Inference} can be seen both as a sub-field of
Information Theory and of general Statistical Inference \cite{03Mac}.
A typical problem in this field is: given
observed noisy data and known statistical model of a noisy
communication channel (transition probability), as well as a {\emph{prior
distribution}} for the input (a pre-image),
find the most likely pre-image, or  compute the {\emph{a posteriori}} marginal
probability for some part of the pre-image.

This field is also deeply related to
  {\it Combinatorial
Optimization},  which is a branch of optimization in Computer
Science, related to operations research, algorithm theory and
complexity theory \cite{98PS}. A typical problem in {\it
Combinatorial Optimization} is: solve, approximate or count (exactly
or approximately) instances of problems by exploring the
exponentially large space of solutions. In many
emerging applications (in magnetic and optical recording,
micro-fabrication, chip design, computer vision, network routing and
logistics), the data are structured in a two-dimensional grid (array).
Moreover, data associated with an element of the grid are often
binary and correlations imposed by the problem are local, so that
only nearest neighbors on the grid are correlated. Such problems are
typically stated in terms of binary statistical models on planar
graphs.

In this paper, we discuss a generic problem of Bayesian inference
defined on a planar graph. We focus on the problem of weighted
counting, or (from the perspective of statistical physics) we aim to
calculate the partition function of an underlying statistical model.
As the seminal work of  Onsager \cite{44Ons} on the two-dimensional
Ising model and its combinatorial interpretation by Kac and Ward
\cite{52KW} have shown, the planarity constraint dramatically
simplifies statistical calculations. By contrast, three-dimensional
statistical models are much more challenging, and no exact results
are known.

Building on the work of physicists,  specifically on results of
Fisher \cite{61Fis,66Fis} and Kasteleyn \cite{61Kas,63Kas}, Barahona
\cite{82Bar} has shown  that calculating the partition function of
the spin glass Ising model on an arbitrary planar graph is
{\emph{easy}}, as the number of operations required to evaluate the
partition function scales algebraically, $O(N^3)$, with the size of
the system. To prove this, the partition function of the spin-glass
Ising model was reduced to a dimer model on an auxiliary graph, and
the partition function was expressed as the Pfaffian of a
skew-symmetric matrix defined on the graph. The polynomial algorithm
was later used in simulations of spin glasses \cite{93SK}. However,
Barahona also added a grain of salt to the exciting positive result,
showing that generic planar binary problem is difficult \cite{82Bar,
Jerrum}. Specifically, evaluating two-dimensional spin glass Ising
model in a magnetic field is NP-hard, i.e. it is a task of likely
exponential complexity.

When an exact computational algorithm of polynomial complexity is
not available, efficient  approximations become relevant. Typically,
the approximation is built around a tractable case. One such
approximate algorithm built around the Fisher-Kasteleyn Pfaffian formula
was recently suggested by Globerson and Jaakkola in \cite{06GJ}. Although
this approximation (coined ``planar-graph decomposition")
gives a provable upper bound for the partition function for some
special graphical models, it constitutes just heuristics, i.e. it suffers
from lack of error-control and the inability of
gradual error-reduction.

Controlling errors in approximate evaluations of the partition function of
a graphical model is generally difficult. However, one recent
approach, developed by two of us  and called Loop Calculus
\cite{06CCa,06CCb}, offers a new method. Loop Calculus allows to
express explicitly the partition function of a general statistical inference problem via
an expansion (the Loop Series), where each term is explicitly expressed via a
solution of the Belief Propagation \cite{63Gal,68Gal,88Pea}, or Bethe-Peierls
\cite{82Bax, 35Bet,36Pei} (BP) equations. This brought new
significance to the BP concept, which previously was seen as
just heuristics.

The BP equations are tractable for any graph; generally, the number of terms in the Loop Series is
exponentially large, so direct re-summation is not feasible.
However, since any individual term in the series can be evaluated
explicitly (once the BP solution is known), the Loop Series
representation offers a possibility for correcting the bare BP
approximation perturbatively, accounting for loop contributions one
after another sequentially. This scheme was shown to work well in
improving BP decoding of Low-Density Parity Check codes in the
error-floor regime,  where the number of important loop
contributions to the Loop Series is (experimentally) small,  and the
most important loop contributions (comparable by absolute value to
the bare BP one) have a simple, single-connected structure
\cite{06CCc,07Che}. In spite of this progress, the question
remained: what to do with other truly difficult cases when the
number of important loop corrections is not small, and when the
important corrections are not necessarily single-connected? In
general, we still do not know how to answer these questions,  while
a partial answer for the important class of planar models is
provided in this paper.

\subsection{Brief Description of Our Results}

In this manuscript we show that, for any graph (planar or not), the
partial sum of the loop series over single-connected loops reduces
to evaluation of the full partition function of an auxiliary dimer-matching
model on an extended, regular degree-3 graph.
Weights of dimers calculated on the extended
graph are expressed explicitly via solution of the respective BP
equations.   The dimer weights can be positive or negative. In
general, summing the single-connected partition is not tractable.
However, in the planar case, it reduces (through manipulations
reminiscent of the Fisher-Kasteleyn transformations) to a Pfaffian
defined on the extended graph,  which is also planar by
construction.
Thus,  we find a big class of planar graphical models which
are computationally tractable by reduction (via a BP/gauge transformation) to
a loop series including only single-connected loops, and summable into a Pfaffian.
Moreover,  we find that the partition function of the
entire Loop Series is generally reducible to a weighted Pfaffians series,
where each higher-order Pfaffian is associated with a sum of dimer configurations
on a modified subgraph of the original graph.  Each term
in the Pfaffian series is computationally tractable via the Belief Propagation
solution on the original graph.

The material in the manuscript is organized as follows. A formal definition of the
model is given in Section \ref{subsec:Model} and a brief description of
Loop Calculus \cite{06CCa,06CCb} forms Section
\ref{subsec:LoopCalc}. Some introductory material on the graphical
transformations is also given in \ref{sec:Graph}. Section
\ref{sec:SingleLoops} is devoted to re-summation of the
single-connected loops in the Loop Series (we called it single
connected partition). Section \ref{subsec:ToDimer} introduces
graphical transformation from the original graph ${\cal G}$ to the
extended graph ${\cal G}_e$,  reminiscent the Fisher transformation
\cite{61Fis,66Fis}. This allows to restate the single-connected
loop partition of the Loop Series on the original graph in terms of
a sum over dimer configurations on the extended graph. Subsection
\ref{subsec:Pfaff} adapts the Kasteleyn transformation
\cite{61Kas,63Kas} to our case, thus expressing the partition
function of the single-connected series as a Pfaffian of a matrix
defined on the extended graph. Section \ref{sec:Easy} describes a
set of graphical models reducible under  Belief Propagation gauge
(transformation) to a Loop Series which is computationally
tractable. Section \ref{sec:Full} describes the representation of
the Loop Series for planar graphs in terms of the Pfaffian Series,
where each Pfaffian sums dimer matchings on a graph extended from a
subgraph of ${\cal G}$, with the later correspondent to exclusion of
an even set of vertices from ${\cal G}$. Grassmann
representations, as well as fermionic models
are discussed in Section \ref{sec:Grass}:
 a general set of Grassmann models on
super-spaces is given in Section \ref{subsec:GrassGen}, while
Section \ref{subsec:comments-fermions} addresses the
relation between binary models and integrable hierarchies.
A  brief  list of future research topics  is given in Section \ref{sec:Con}.

\subsection{Vertex-function Model}
\label{subsec:Model}

We introduce an undirected graph ${\cal G}=({\cal V},{\cal E})$
consisting of vertices ${\cal V}=(a=1,\cdots,N)$ and edges ${\cal
E}$. This  study focuses mainly on planar graphs, like those
emerging in communication or logistics networks connecting or
relating nearest neighbors on a 2d mesh or terrain. However,  the
material discussed in the present and the following Subsections is general,
and applies to any graph, planar or not. A binary variable,
$\sigma_{ab}=\pm 1$, which we will also be calling a spin, is
associated with any edge $(a,b)\in{\cal E}$. The graphical model is
defined in terms of the probability function
\begin{eqnarray}
p(\vec{\sigma})=Z^{-1}\prod_{a\in{\cal V}}f_a(\vec{\sigma}_a),
\label{P_sigma}
\end{eqnarray}
for a spin configuration $\vec{\sigma}\equiv \{ \sigma_{ab}=\pm 1| \forall
 (a,b)\in{\cal E} \}$. In (\ref{P_sigma}),
$\vec{\sigma}_a=(\sigma_{ab}| \forall b,\mbox{ s.t. }(a,b)\in{\cal
E})$ is the vector built from all edge variables associated with the
given vertex $a$. $f_a$'s are positive and otherwise we will assume no
restrictions on the factor functions. $Z$  is
the normalization factor, the so-called partition function of the
graphical model.

We refer to (\ref{P_sigma}) as ``vertex-function" models, according to
statistical physics notation \cite{82Bax}.
In the information theory, they  are
known as Forney-style graphical models
\cite{01For,01Loe}.

We will assume in the following that the degree of connectivity of any
vertex in the graph is three. Note that this is not a restrictive
condition,  as the $n$-th order vertices, correspondent to $n$-spin
interactions with $n>3$, can always be represented in terms of a
product of triplet terms. Then the $n$-th degree vertex can be
transformed into a planar graph consisting of degree three vertices.
We discuss transformations to the triplets, in general but also on
some examples (Ising Model and Parity Check Decoding of a linear
code), in \ref{sec:Graph}.

\subsection{Loop Calculus}
\label{subsec:LoopCalc}

Loop Calculus \cite{06CCa,06CCb} gives an
explicit expression for  $Z$ through the Loop Series:
\begin{eqnarray}
 && Z=Z_0 \cdot z,\,\, z\equiv \left(1+\sum_{\it C}
 \prod\limits_{a\in{\it C}}\mu_{a,\bar{a}_{\it C}}\right),\,\,
 \mu_{a,\bar{a}_{\it C}}\equiv\frac{\tilde{\mu}_{a,\bar{a}_{\it C}}}
 {\prod\limits_{b\in {\it C}}^{(a,b)\in {\it C}}\sqrt{1-m_{ab}({\it C})}} \label{Zseries}\\
 && m_{ab}=\sum_{\sigma_{ab}}\sigma_{ab}b_{ab}(\sigma_{ab}),\,\,
 \tilde{\mu}_{a,\bar{a}_{\it C}}=\sum_{\vec{\sigma}_a}\prod_{b\in \bar{a}_{\it C}}(\sigma_{ab}-m_{ab})
 b_a(\vec{\sigma}_a),
 \label{mu_ab_a}
\end{eqnarray}
where ${\it C}$ can be any allowed generalized loop on the graph
${\cal G}$,  i.e. ${\it C}$ is a subgraph of ${\cal G}$ which does
not contain any vertices of degree one; $\bar{a}_{\it C}$ is a set of vertices of graph ${\cal G}$
which are also contained in the generalized loop ${\cal C}$
(by construction $\bar{a}$ consists of two or three elements);
and $b_a(\vec{\sigma}_a)$
and $b_{ab}(\sigma_{ab})$ are beliefs associated with vertex $a$ and
edge $(ab)$.  The beliefs are defined via message variables
$\eta_{ab}\neq\eta_{ba}$
\begin{eqnarray}
 && \forall\ (a,b)\in{\cal E}:\quad
 b_{ab}(\sigma_{ab})=\frac{\exp\left((\eta_{ab}+\eta_{ba})\sigma_{ab}\right)}{
 2\cosh\left(\eta_{ab}+\eta_{ba}\right)},\label{bab}\\
 && \forall\ a\in{\cal V}:\quad
 b_a(\vec{\sigma}_a)=\frac{f_a(\vec{\sigma}_a)\exp\left(\sum_b^{(a,b)\in{\cal
 E}}\eta_{ab}\sigma_{ab}\right)}{\sum_{\vec{\sigma}_a}
 f_a(\vec{\sigma}_a)\exp\left(\sum_c^{(a,c)\in{\cal E}}\eta_{ac}\sigma_{ac}\right)},
 \label{ba}
\end{eqnarray}
solving the following system of the Belief
Propagation (BP) equations
\begin{equation}
\!\!\!\!\!\!\!\!\!\!\!\!\forall\ (a,b)\in{\cal E}:\quad
 \sum_{\vec{\sigma}_a}
 f_a(\vec{\sigma}_a)\exp\left(\sum_b^{(a,b)\in{\cal
 E}}\eta_{ab}\sigma_{ab}\right)\left(\sigma_{ab}-\tanh\left(\eta_{ab}+\eta_{ba}\right)\right)=0.
 \label{bb}
\end{equation}
The bare (BP) partition function $Z_0$ in Eq.~(\ref{Zseries}) has
the following expression in terms of the message variables:
\begin{equation}
 Z_0=\frac{\prod_a\sum_{\vec{\sigma}_a\in{\cal V}}f_a\left(\vec{\sigma}_a\right)
 \exp\left(\sum_{(a,b)\in{\cal E}} \eta_{ab}\sigma_{ab}\right)}
 {\prod_{(a,b)\in{\cal E}}\left [ 2\cosh\left ( \eta_{ab}+\eta_{ba}\right)\right ]}.\label{Z0}
\end{equation}

BP equations (\ref{bb}) are interpreted as
conditions on the gauge transformations, leaving the partition
function of the model invariant. These equations may allow multiple solutions,  related to each
other via respective gauge transformations. The multiple solutions
correspond to multiple extrema of the Bethe Free Energy and
Loop Series can be constructed around any of the BP solutions.  \footnote{See \cite{06CCa,06CCb,06CCc} for a detailed discussion of this and other related features of BP equations as gauge fixing conditions.}

\section{Re-summation of the Single-connected Partition}
\label{sec:SingleLoops}

In the following we will show how to re-sum a part of the Loop
Series accounting for all the single-connected loops,  i.e.
subgraphs of ${\cal G}$ with all vertices of degree two
\begin{eqnarray}
Z_s=Z_0 \cdot z_s,\quad z_s=1+\sum_{{\it C}\in{\cal G}}^{\forall a\in {\it
C},\ |\delta(a)|_{\it C}=2}r_{\it C}, \label{sub_graph}
\end{eqnarray}
where $|\delta(a)|_{\it C}$ stands for the number of neighbors of
$a$ within ${\it C}$. The evaluation will consist of the following
two steps:
\begin{itemize}

\item[A)]  Show that $z_s$ is equal to
the partition function of the dimer-matching model on an auxiliary
graph, ${\cal G}_e$.  The graph will be constructed from the
original ${\cal G}$ by a transformation reminiscent of the Fisher's
trick, introduced in \cite{61Fis,66Fis,82Bar} to streamline
reduction of Ising model to the dimer-matching model;

\item[B)] Use the Pfaffian formula of Kasteleyn \cite{61Kas,63Kas,82Bar} to
reduce $z_s$ to a Pfaffian of a skew-symmetric matrix defined on
${\cal G}_e$. Note that complexity of the Pfaffian evaluation is
$N^3$,  where $N$ is the size of ${\cal G}$.

\end{itemize}

\noindent Note: while  A) is valid for any graphical model, B) applies only to the planar case.

\subsection{Transformation to Dimer Matching Problem}
\label{subsec:ToDimer}

\begin{figure}[t]
\begin{center}
    \includegraphics[width=6cm]{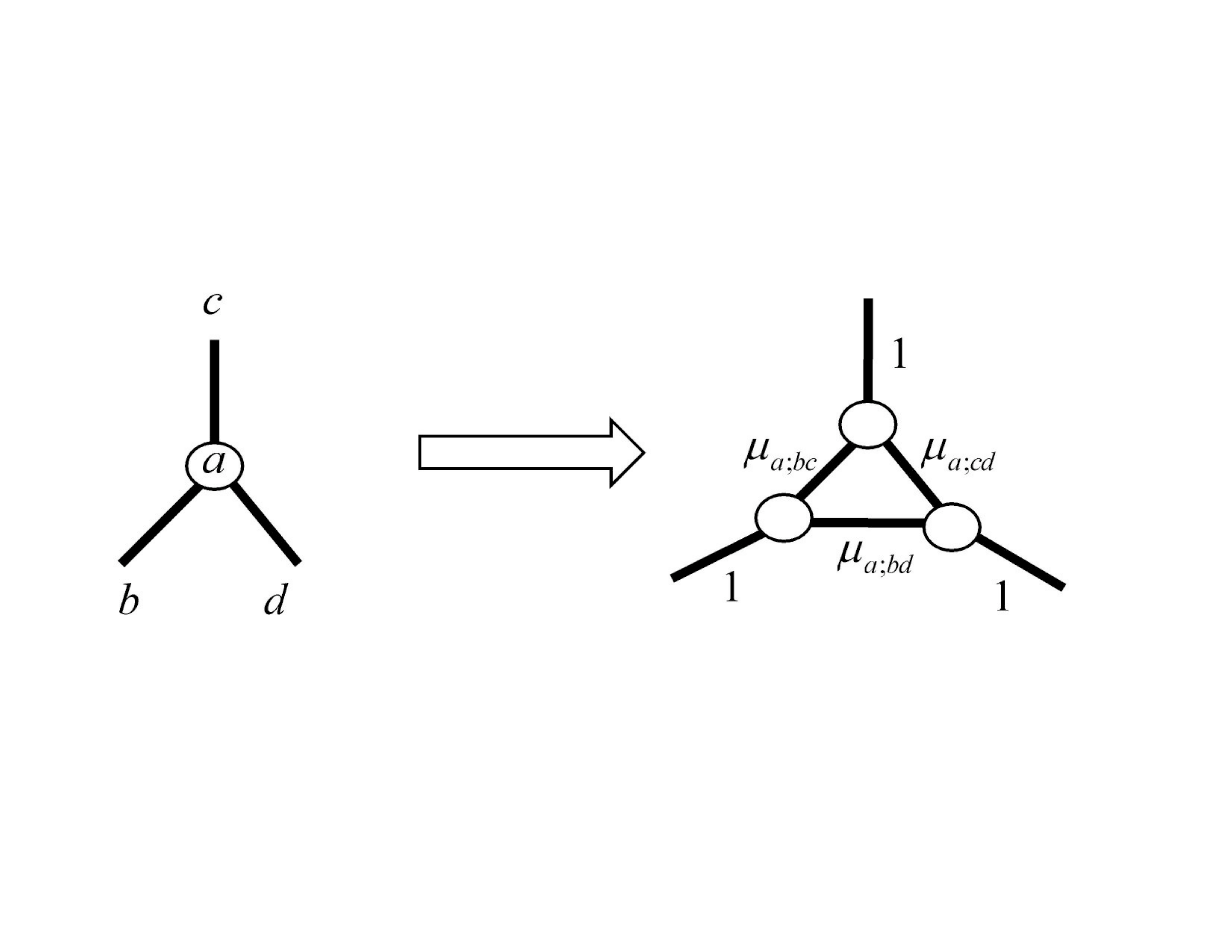}
    \includegraphics[width=6cm]{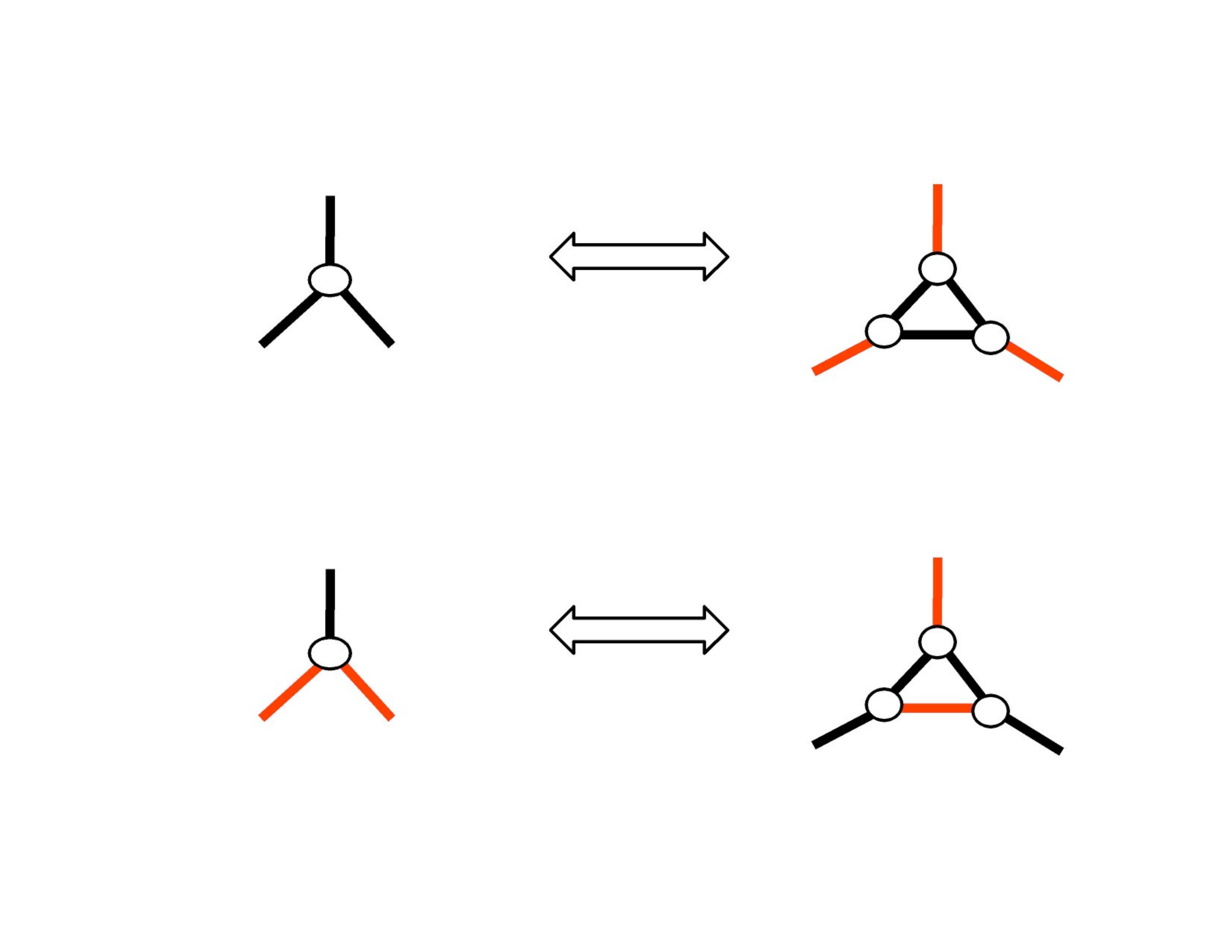}
    \caption{Left panel: Transformation from a vertex of ${\cal G}$ to
    respective three-vertex of the extended
    graph ${\cal G}_e$. Right panel: maps from the colorings of a vertex
    of ${\cal G}$ to coloring of the respective $3$-vertex of ${\cal G}_e$.
    Notice,  that the coloring of the external edges of
    ${\cal G}_e$ are reversed in comparison with the coloring of original edges on ${\cal G}$.}
    \label{trans}
    \end{center}
\end{figure}

\begin{figure}
\begin{center}
    \includegraphics[width=6cm]{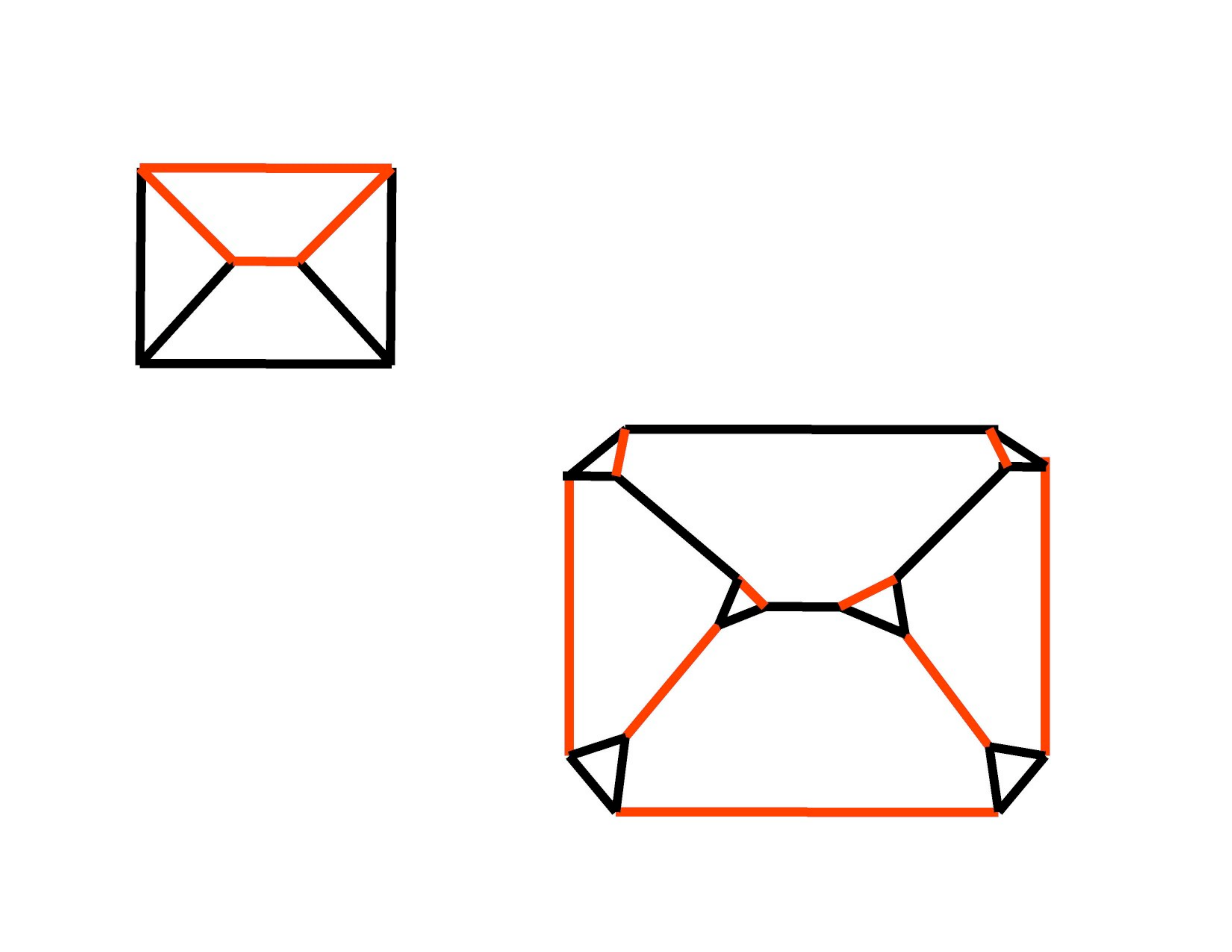}
    \caption{Example of ${\cal G}$ (upper left) to ${\cal G}_e$ (lower right) map.
    Single connected loop of ${\cal G}$ (shown in red) is in one-to-one correspondence with
    a valid dimer matching of ${\cal G}_e$, where dimers are also shown in red.}
    \label{trans_example}
    \end{center}
\end{figure}

Following the construction of Fisher \cite{61Fis,66Fis}, we expand
each vertex of ${\cal G}$ into a three-vertex of the extended graph
${\cal G}_e$, according to the scheme shown in the left panel of
Figure~1.
Consider a vertex $a$ of ${\cal G}$ and assume that $b,c,d$ are
three neighbors of $a$ on ${\cal G}$. For each vertex $a$, there are three
$\mu_{a;\bar{a}_{\it C}}$ contributions
of degree two within a generalized loop ${\it C}$, i.e. with $|\delta \bar{a}_{\it C}|=2$,
which can possibly
contribute to the single-connected partition $r_s$:
$\mu_{a;bc},\mu_{a;bd},\mu_{a;cd}$. We associate the three weights with internal edges of
the respective three-vertex of ${\cal G}_e$, while the weights of all the external edges of the
three-vertex are equal to unity.
Then any coloring of the original graph, marking  a single connected loop
of ${\cal G}$, is in the one-to-one correspondence to a dimer-matching (which we also call
coloring) of ${\cal G}_e$.
The weights and coloring assignments are illustrated on an example at the left panel of Figure~1.
An example of transformation mapping a single-connected-loop on ${\cal G}$
respective dimer on ${\cal G}_e$ is shown in Figure~\ref{trans_example}.

This map from the single-connected loops to dimers leads to the following representation for
the single-connected partition $z_s$
\begin{eqnarray}
 z_s=\sum_{\vec{\pi}}\prod_{(a,b)\in{\cal G}_e}\left(\mu_{ab}\right)^{\pi_{ab}}
 \prod_a \delta\left(\sum_b^{(a,b)\in{\cal G}_e}\pi_{ab},1\right),
 \label{zsd}
\end{eqnarray}
where the dimer-weights on ${\cal G}_e$ are defined according to the simple rules
explained in the previous paragraph.
One finds that the right hand side of (\ref{zsd}) is nothing but the partition function
of a dimer-matching problem on ${\cal G}_e$.

\subsection{Pfaffian Expression for the Partition Function}
\label{subsec:Pfaff}

\begin{figure}[t]
\begin{center}
    \includegraphics[width=6cm]{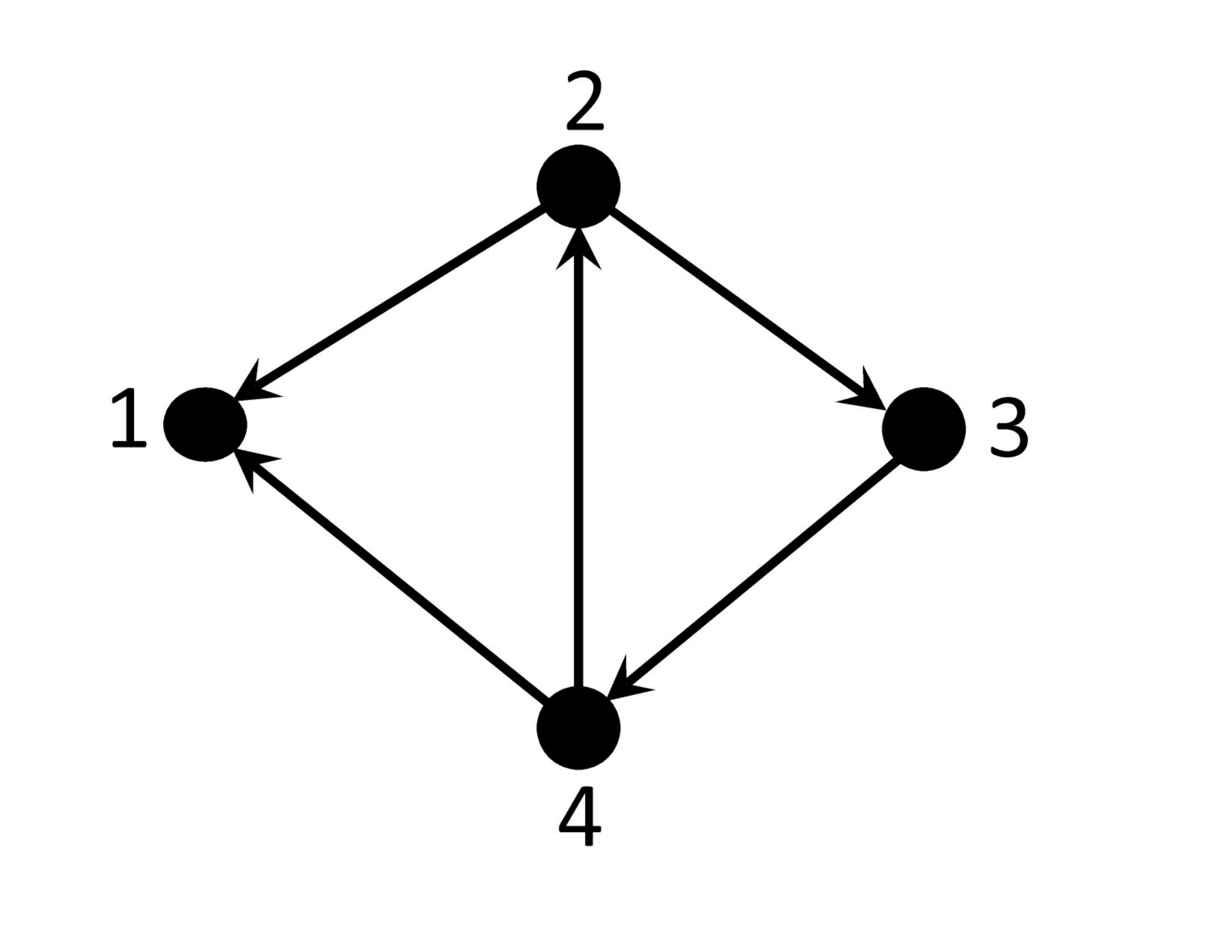}
    \includegraphics[width=6cm]{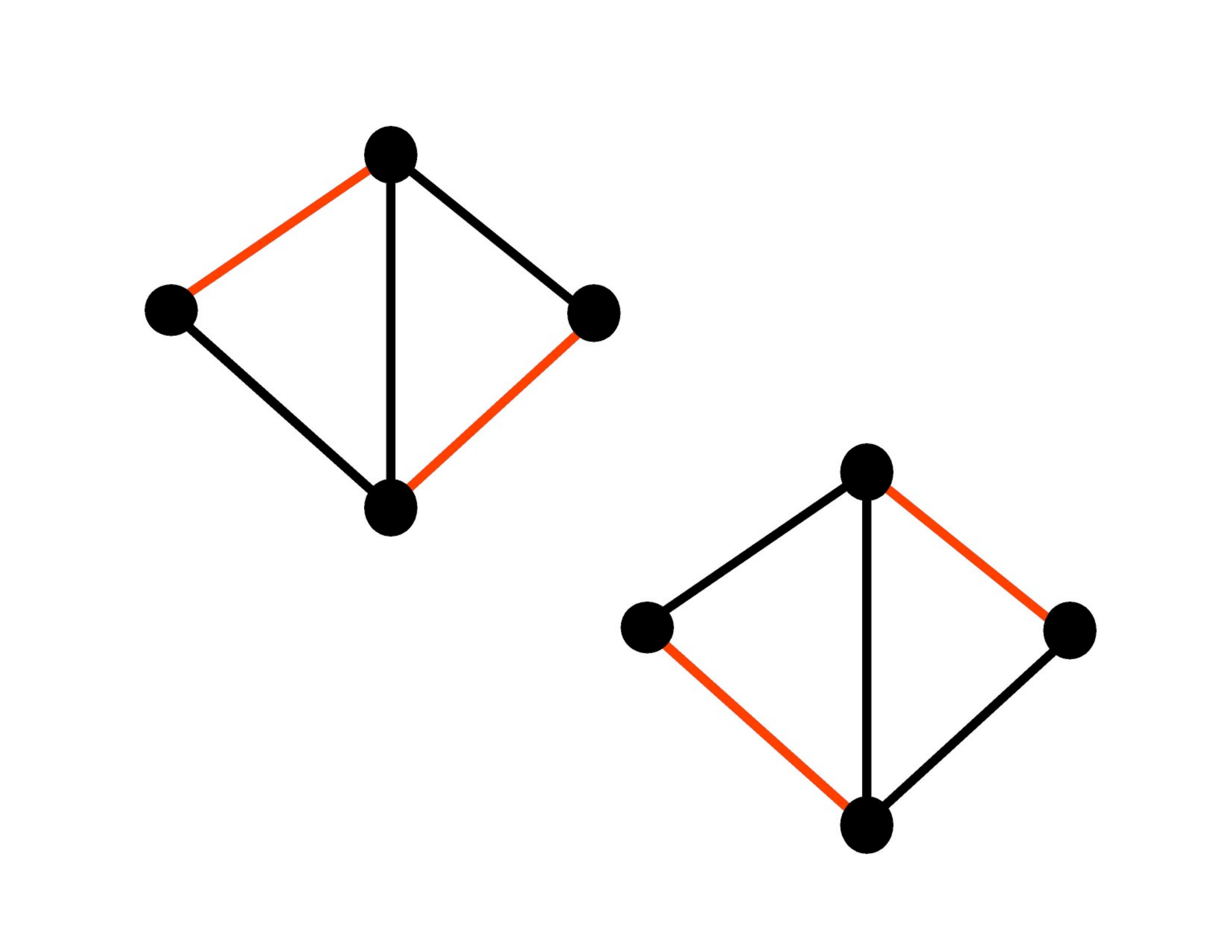}
    \caption{$\vec{p}$ orientation (left panel) and respective dimer (matching) configurations
    (right panel) correspondent to  example of ${\cal G}_e$ described by Eq.~(\ref{Pfaf-eq}).}
    \label{Pfaf-fig}
     \end{center}
\end{figure}

Kasteleyn has shown in \cite{61Kas,63Kas} (see also \cite{82Bar})
that $z_s$ is equal to a Pfaffian (the square root of determinant)
of a skew-symmetric matrix $\hat{A}=-\hat{A}^{t}$ of
size $N_a\times N_a$,  where $N_a$ is the number of vertices in
${\cal G}_a$.  Each element of the matrix with $a>b$ (ordering is
arbitrary,  but it is fixed once and forever) is
$A_{ab}=p_{ab}z_{ab}$, where $p_{ab}=\pm 1$. There are many possible
choices of $\vec{p}=(\pi_{ab}=\pm 1|(a,b)\in{\cal G}_e)$  which
guarantee the Pfaffian relation: $z_s=\sqrt{\mbox{det}\hat{A}}$. A
simple constructive way of choosing such a valid $\vec{p}$ is to
relate it to orientation of edges in a directed version of ${\cal
G}_a$, built according to the following ``odd-face" rule: number of
clockwise-oriented segments of any internal face  of ${\cal G}_e$ should be negative. \footnote{Except,
possibly, the external face.}
Example of a valid orientation is shown in Figure~\ref{trans_example} and the
respective expressions are
\begin{equation}
 \!\!\!\!\!\!\!\!\!\!\!\!\!\!\!
 z_{s;{\it example}}=\mu_{12}\mu_{34}+\mu_{14}\mu_{23}=
 \sqrt{\mbox{Det}\hat{A}},\quad
 \hat{A}=\left(\begin{array}{cccc} 0 & -\mu_{12} & 0 &
 -\mu_{14}\\ \mu_{12} & 0 & \mu_{23} & -\mu_{24}\\
 0 & -\mu_{23} & 0 & \mu_{34}\\
 \mu_{14} & \mu_{24} & -\mu_{34} & 0\end{array}\right)
 \label{Pfaf-eq}
\end{equation}

Since calculating the determinant requires $\sim N_a^3$ operations,
one finds that re-summation of all the single-connected loops in the
Loop Series expression for the partition function of a planar
graphical model can be done efficiently in $O(N_a^3)$ steps.

\section{Tractable Problems Reducible to Single-Connected Partition}
\label{sec:Easy}

In the case of a general vertex-function graphical model, the
BP-gauge transformations, described by the set of BP equations
(\ref{bb}), result in exact cancelation in the Loop Series of all
the subgraphs containing at least one vertex of degree one within
the subgraph. Thus, for the graph with all vertices of degree three,
any vertex contributing a generalized loop (subgraph) should be of
degree two or three within the subgraph. As shown in the previous
Section, if one ignores generalized loops with vertices of degree
three and the original graph is planar, the resulting sub-series
(single-connected partition) is computationally tractable, i.e. the
number of operations required to evaluate the single-connected
partition is cubic in the system size (not exponential !).

In this Section we discuss the class of planar models whose
Loop Series do not contain any generalized loops with
vertices of degree three. According to Section
\ref{sec:SingleLoops}, these models are tractable.

Indeed, it is known that BP Eqs.~(\ref{bb}) have at least one
solution for the set of messages $\{\eta\}$ on any graph and for any
factor functions defined on the vertices of the graph. The
aforementioned requirement for the generalized loop not to contain
any vertex of degree three translates
into the following set of additional equations
\begin{equation}
 \!\!\!\!\!\!\!\!\forall\ a\in{\cal G}:\quad
 \sum_{\vec{\sigma}_a}
 f_a(\vec{\sigma}_a)\prod_{b}^{(a,b)\in{\cal E}}\left(\exp\left(\eta_{ab}\sigma_{ab}\right)
 \left(\sigma_{ab}-\tanh\left(\eta_{ab}+\eta_{ba}\right)\right)\right)=0.
 \label{b3}
\end{equation}
Considered together, the set of Eqs.~(\ref{bb},\ref{b3}) is
overdefined,  i.e. it cannot be solved in terms of $\eta$ variables
for any values of the factor functions.   However, if one allows
flexibility in the factor functions,  and, in fact, considers
Eqs.~(\ref{bb},\ref{b3}) as a set of conditions on both the messages
$\{\eta\}$ and the factor functions $\{f\}$, one arrives at a big
set of possible solutions.

Therefore,  Eqs.~(\ref{bb},\ref{b3}) define a big set of models
reducible via BP transformations to a tractable Loop Series
consisting only of single connected loops.

Moreover,  the relations we established may be reversed. One
may start from an arbitrary Loop Series consisting of only single
connected loops,  apply an arbitrary gauge transformation leaving
the Loop Series invariant (these transformations are
not necessarily of BP type),  and arrive at a graphical model
with some set of factor functions.  At first sight, the  resulting
graphical model might not look tractable, but it actually is,
by construction.

\section{Loop Series as a Pfaffian Series}
\label{sec:Full}

Let us notice that the general planar problem (e.g. spin
glass in a magnetic field) is NP-hard
\cite{82Bar}, and it is thus not surprising that full re-summation
does not allow expression in terms of a single Pfaffian
(or a determinant).

On the other hand,  we already found that a part of the Loop Series,  specifically its single-connected
partition, reduces to a computationally tractable Pfaffian. This suggests to represent the full
Loop Series as a sum over terms, each representing a set of triplets (fully colored vertices of degree
tree on ${\cal G}$):
\begin{equation}
z=\sum_{\Psi} z_\Psi\prod_{a\in\Psi}^{|\bar{a}|=3}\mu_{a;\bar{a}},\label{ser_trip}
\end{equation}
where $\Psi$ is either the  empty set or any set of even nodes on ${\cal G}$; $\mu_{a;\bar{a}}=\mu_{a;bcd}$
are the weights from
Eq.~(\ref{Zseries}) associated with the triplet $(a;b,c,d)$,  such that $(a,b),(a,c),(a,d)\in{\cal E}$;
and $z_\Psi$ is the sum over all generalized loops
(proper Loop Series colorings, i.e. subgraphs) of ${\cal G}$ such that all nodes of $\Psi$ are fully colored
(all edges adjusted to the nodes belong to the generalized loop),
while any other vertices of ${\cal G}$ are not colored or only partially colored.
Thus,  the first term in Eq.~(\ref{ser_trip}), where $\Psi$ is the empty set, represents
the single-connected partition,  $z_s$.

We show here that not only the first term in Eq.~(\ref{ser_trip}), associated with $\Psi=\varnothing$,
but any  term $z_\Psi$ in Eq.~(\ref{ser_trip}) is computationally tractable, being equal to a Pfaffian of
a matrix defined on ${\cal G}_e$.

Indeed, it is straightforward to verify that the generalized loops associated with the given
set of triplets (fully colored vertices) from the set $\Psi$ are in
one-to-one correspondence with the set
of dimer matchings on ${\cal G}_{e;\Psi}$, which is a subgraph of ${\cal G}_e$
with all $3$-vertices correspondent to $\Psi$, and external edges connected to the vertices,
completely removed. Notice that some vertices of  ${\cal G}_{e;\Psi}$ are of degree two.
(These are vertices neighboring the removed triplets of $\Psi$.)

\begin{figure}
\begin{center}
    \includegraphics[width=6cm]{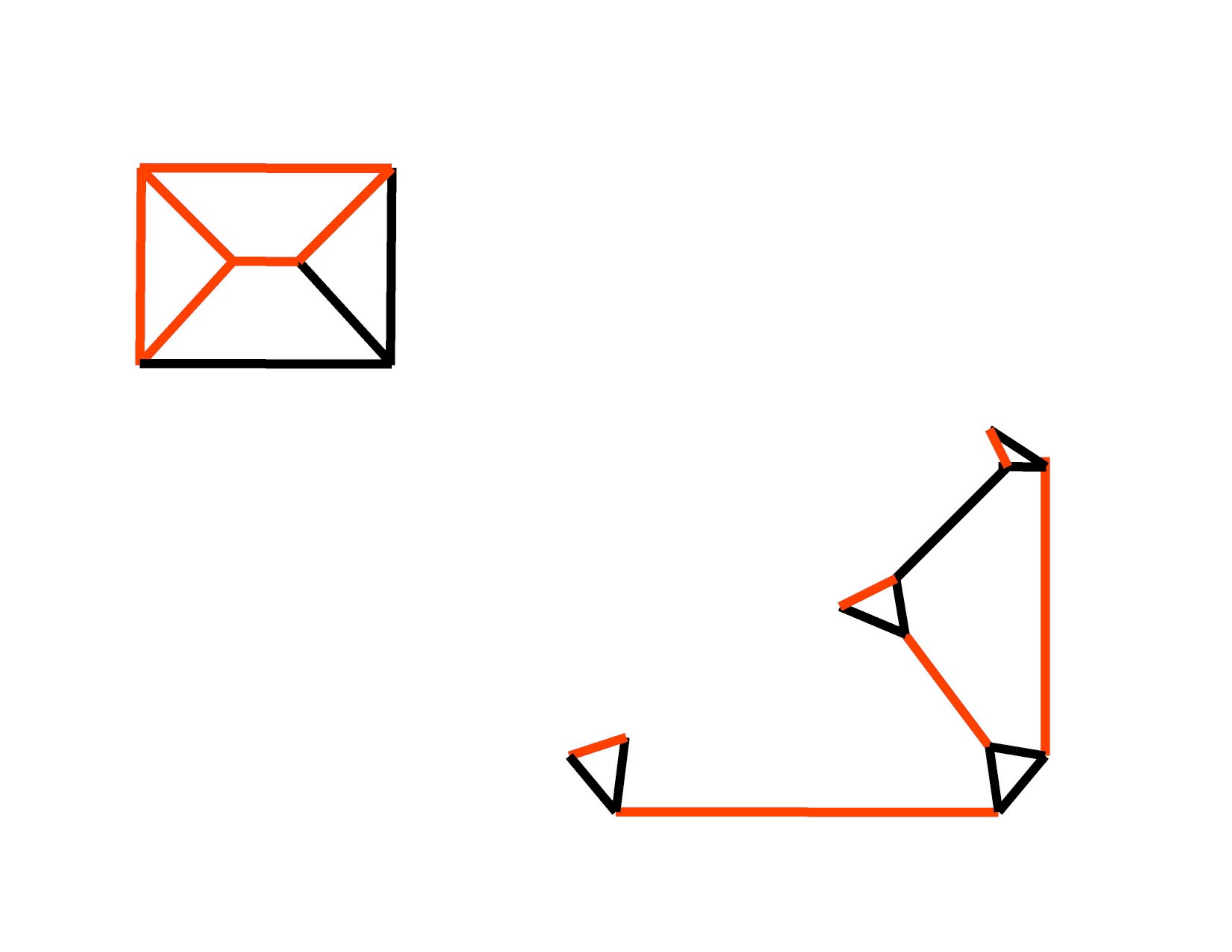}
    \includegraphics[width=6cm]{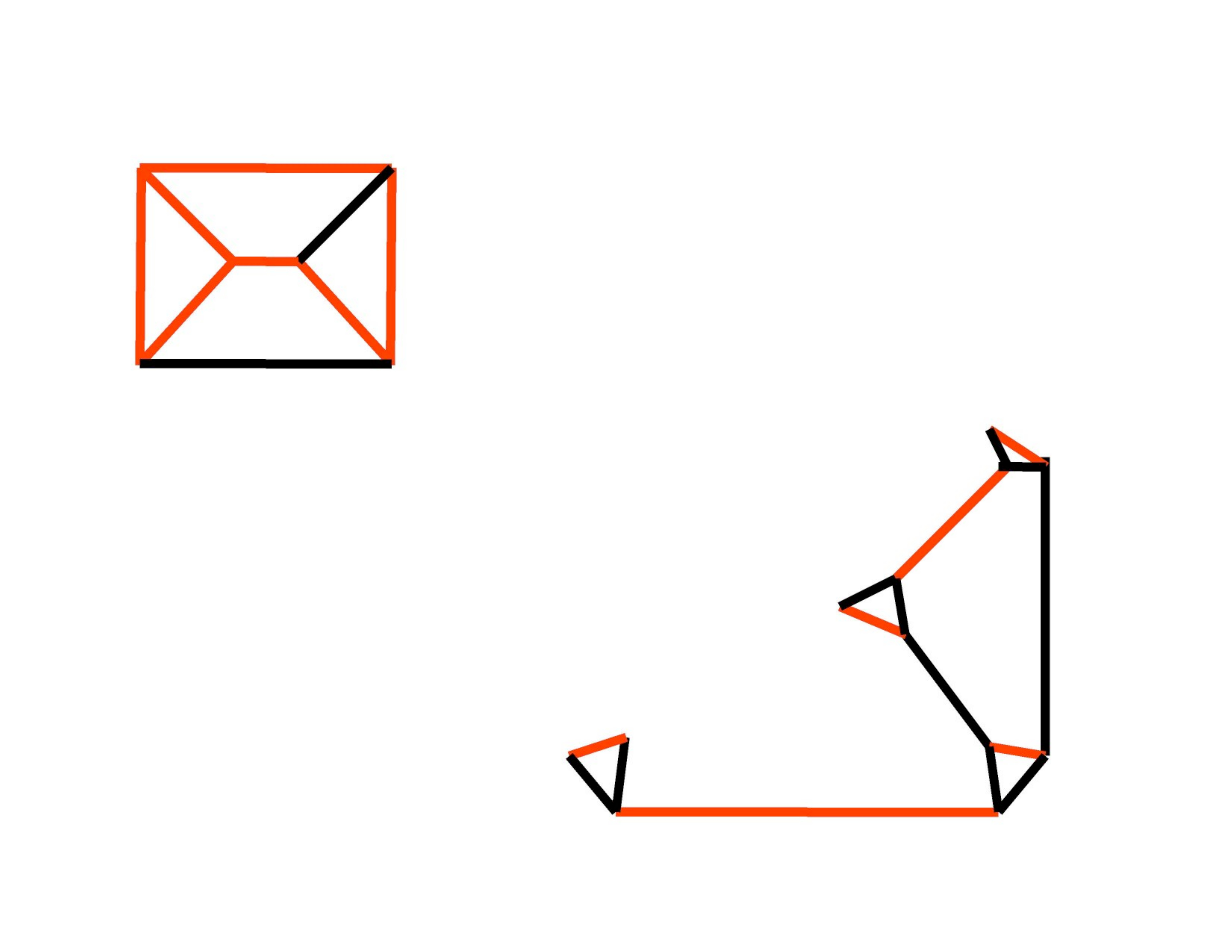}
    \caption{Two generalized loops (shown on the top) of an exemplary ${\cal G}$ correspondent to the same
    configuration of triplets $G$, $|G|=2$, and their respective dimer configurations on
    ${\cal G}_{e;\Psi}$ (shown on the bottom).}
    \label{fig:Psi}
     \end{center}
\end{figure}

An example of a  ${\cal G}_{e;\Psi}$
construction is given in Figure~\ref{fig:Psi}.
One associates weights to the edges of ${\cal G}_{e;\Psi}$ in exactly the same way
as for the single-connected partition: the weights of all the external edges of $3$-vertices
of  ${\cal G}_{e;\Psi}$ are equal to unity,  while the internal edges are associated with the
respective values $\mu_{a;bc}$, defined in Eq.~(\ref{mu_ab_a}).

For any of ${\cal G}_{e;\Psi}$ one constructs the skew-symmetric $\hat{A}_\Psi$ matrix
according to the Kasteleyn rule for the dimer-matching model described in Section \ref{subsec:ToDimer}.
As before,  the dimensionality of the matrix is $|{\cal G}_{e;\Psi}|\times |{\cal G}_{e;\Psi}|$ and
each element of the matrix is the product of the respective dimer weight and orientation sign.
Notice  that the choice of signs for
the elements of $\hat{A}_\Psi$ depends on the set of ``excluded" triplets $\Psi$,
and thus $\hat{A}_\Psi$ is not simply a minor of the original
matrix $\hat{A}$,  the one corresponding to the single-connected partition (without exclusion). Thus,
\begin{eqnarray}
z_{\Psi}=\mbox{Pf}\left(\hat{A}_{\Psi}\right)=
\sqrt{\mbox{Det}\left(\hat{A}_{\Psi}\right)}.
\label{zdm2}
\end{eqnarray}
Eqs.~(\ref{ser_trip},\ref{zdm2}) describe the  Pfaffian series representation for the
Loop Series of the planar problem.

\section{Fermion Representation and Models}
\label{sec:Grass}

Any Pfaffian in Eq.(\ref{zdm2}) allows a compact representation in
terms of Grassmann variables \cite{87Ber}. Indeed, let us associate a
Grassmann (anti-commuting or fermionic) variable $\theta_a$ with
each vertex of ${\cal G}_e$. The Grassmann variables satisfy
\begin{eqnarray}
\forall (a,b)\in{\cal G}_e:\quad
\theta_a\theta_b+\theta_b\theta_a=0, \label{Grass}
\end{eqnarray}
and commute with ordinary $c$-numbers. One also introduces the
Berezin integration rules over the Grassmann variables
\begin{eqnarray}
\int d\theta=0,\quad\int \theta d\theta=1.\label{Ber}
\end{eqnarray}
This translates into the following rule of Gaussian integration over
the Grassmann variables:
\begin{eqnarray}
\int
\exp\left(-\frac{1}{2}\vec{\theta}^{t}\hat{A}\vec{\theta}\right)d\vec{\theta}=\mbox{Pf}(\hat{A})=
\sqrt{\mbox{det}(\hat{A})}, \label{BerGauss}
\end{eqnarray}
where $\vec{\theta}$ is the vector of the Grassmann variables over
the entire graph, $\vec{\theta}=\left(\theta_i|i\in{\cal
G}_a\right)$ and $\hat{A}$ is an arbitrary skew-symmetric matrix on
the graph. For example, applying this formula to the first term of the Pfaffian
series (\ref{ser_trip}) one derives
\begin{eqnarray}
z_{\vec{0}}=\int
\exp\left(-\frac{1}{2}\vec{\theta}^{t}\hat{A}\vec{\theta}\right)d\vec{\theta}.
\label{Grassmann0}
\end{eqnarray}
In general,  any term in the Pfaffian series of Eq.~(\ref{ser_trip}) can be represented
as a Gaussian Grassmann integrable,  however with different Gaussian kernels,  not reducible simply to
minors of $\hat{A}$.

\subsection{Graphical Models on Super-Spaces}
\label{subsec:GrassGen}

In this Subsection we first consider graphical models on spaces
generalizing the $2$-point (binary) set to super-spaces containing
commuting and anti-commuting parts. The models will be defined on
arbitrary (non necessarily planar) graphs. Then, we return to
the simple example (\ref{Grassmann0}) of pure dimer model with the
Grassmann (anticommuting) variables defined on vertices of ${\cal
G}_e$, to see that the model can be restated as the vertex-function
Grassmann model on the original graph ${\cal G}$.

The general class of vertex-function models can be introduced
as follows. For our graph, ${\cal G}$, consider a set of spaces
$\{M_{a\alpha}|a\in\alpha\}$, i.e., we associate a space with any
edge, $\alpha$, together with a vertex, $a$, that belongs to the
edge. For simplicity we assume the spaces to be identical, i.e.,
$M_{a\alpha}\cong M$ for all $a\in\alpha$. The basic variables are
$\sigma_{a\alpha}\in M_{a\alpha}$. We also introduce the notation (all products  below are cartesian)
\begin{eqnarray}
\label{define-spaces} M_{a}=\prod_{\alpha\ni a}M_{a\alpha}, \;\;
M_{\alpha}=\prod_{a\in\alpha}M_{a\alpha}, \;\; {\cal
M}=\prod_{a\alpha}^{a\in\alpha}M_{a\alpha}=\prod_{a}M_{a}=\prod_{\alpha}M_{\alpha};
\;\; \\
\vec{\sigma}_{a}\in M_{a}, \;\; \vec{\sigma}_{\alpha}\in
M_{\alpha}, \;\; \vec{\sigma}\in{\cal M}.
\end{eqnarray}
Note that any $M_{\alpha}$ is a two-component cartesian product. The
vertex-function model is determined by a set of vertex functions
$f_{a}(\vec{\sigma}_{a})$ defined on $M_{a}$ and a set of
integration measures $d\mu_{\alpha}(\vec{\sigma}_{\alpha})$ on
$M_{\alpha}$. The model partition function is
\begin{eqnarray}
\label{define-Z-continuous} Z=\int_{{\cal
M}}\prod_{\alpha}d\mu_{\alpha}(\vec{\sigma}_{\alpha})\prod_{a}f_{a}(\vec{\sigma}_a).
\end{eqnarray}
For the particular case when measures have supports restricted to
the diagonals $M\cong \Delta_{\alpha}\subset M_{\alpha}\cong M\times
M$, i.e. ${\rm supp}\mu_{\alpha}\subset\Delta_{\alpha}$, we can consider
the basic variables that belong to the diagonals. This corresponds
to a more conventional formulation of the vertex-function models
with the variables residing on edges. Note that the
models introduced allow for loop-tower calculus \cite{07CC}, formulated in
terms of fixing a proper gauge. The BP gauge fixing for a general
vertex-function model described by Eq.~(\ref{define-Z-continuous})
is nothing more than choosing basis sets in the vector spaces (maybe
infinite-dimensional) of functions in $M_{a\alpha}$. A standard
binary model, defined in Eq.~(\ref{P_sigma}), corresponds to the
choice $M=\{0,1\}$ of the basic space to be a $2$-point set. Vertex
models with $q$-ary alphabet, e.g. discussed in \cite{07CC}, are
described by $M=\{0,1,\ldots,q-1\}$. Continuous models are obtained
if $M$ is chosen to be a manifold of dimension $m$. The continuous
case can be extended to the choice of $M$ to be a supermanifold $M$
of dimension $(m_{+},m_{-})$ that contains $m_{-}$ Grassmann
(anticommuting) coordinates and whose substrate $\bar{M}\subset M$
is an $m_{+}$-dimensional manifold. Note that a manifold can be
considered as a supermanifold with zero odd dimension $m_{-}=0$. In
the remainder of this Subsection we will be dealing with an opposite
case of the zero even dimension $m_{+}=0$, specifically with the purely Grassmann case of the
$(0,1)$ supermanifold.

Eq.~(\ref{Grassmann0}) is the partition function of a model stated in terms of Grassmann variables
defined on the vertices of ${\cal G}_e$. The extended graph ${\cal G}_e$ is constructed from the
original graph ${\cal G}$ so that a vertex  of ${\cal G}$ extends into a triangle with three
vertices of degree three (see the left panel of Figure~\ref{trans}). Therefore, the three Grassmann
variables in (\ref{Grassmann0}) are associated with a vertex of ${\cal G}$. Then,
Eq.~(\ref{Grassmann0}) defined on ${\cal G}_e$ allows an obvious reformulation in the
vertex-function form (\ref{define-Z-continuous}) on ${\cal G}$, where $\vec{\sigma}_{a}$ represents
the three Grassmann variables that reside on the vertices of ${\cal G}_{e}$, obtained by expanding
the vertex $a$ of the original graph. The dimer weights for the three edges of ${\cal G}_{e}$
associated with the extended vertex of ${\cal G}$ are encoded in the Gaussian function
$f_{a}(\vec{\sigma}_{a})$. The dimer weight associated with an edge of ${\cal G}_{e}$ that
represents and edge $\alpha$ of the original graph ${\cal G}$ is encoded in the integration measure
$d\mu_{\alpha}(\vec{\sigma}_{\alpha})$.

Also notice that the vertex-function Grassmann model on a planar
graph ${\cal G}$ can be restated as a model on the triangulated
graph,  dual to ${\cal G}$, with complex fermion (Grassmann)
variables associated with the edges of the dual graph and functions
associated with a face (elementary triangle) of the dual graph
(Figure~\ref{fig:Triang} illustrates the duality transformation).
One interesting conclusion here is that the sequence of
transformations discussed above leads us from a special binary model
on a planar graph ${\cal G}$ to a Gaussian fermion (Grassmann) model
on the dual graph, thus representing an instance of the disorder
operator approach of Kadanoff-Ceva \cite{71KC} developed originally
for the Ising model on a square lattice.

\subsection{Comments on Relation to Quantum Algorithms and Integrable Hierarchies}
\label{subsec:comments-fermions}

A mapping of a classical inference problem onto finding an
expectation value in a corresponding quantum model takes on a
natural interpretation as a {\emph{quantum}} algorithm. This can be
tried by using the theory of the infinite Kadomtsev-Petviashvilii (KP)
hierarchy, specifically its fermionic formulation \cite{SMJ}. Consider $1D$
lattice fermions $\psi_{k},\psi_{k}^{*}$ with $k\in{\mathbb Z}$ and
introduce the population $\widehat{n}_{k}=\psi_{k}^{*}\psi_{k}$ and
shift  operators $\widehat{H}_{k}=\sum_{j\in{\mathbb
Z}}\psi_{k+j}^{*}\psi_{j}$. Let $|0\rangle$ denote the standard
many-particle vacuum state where all single-fermion orbitals with
$k\leq 0$ are occupied, and $|W\rangle$ is some uncorrelated (i.e.
represented by a single Slater determinant) many-particle state,
which is sufficiently close to $|0\rangle$. Introducing ${\bm
t}=t_{1},t_{2},\ldots$, $\bar{{\bm
t}}=\bar{t}_{0}, \bar{t}_{1},\bar{t}_{2},\ldots$, and
${\bm\xi}=\ldots,\xi_{-1},\xi_{0},\xi_{1},\ldots$ we consider an
expectation value
\begin{eqnarray}
\label{define-Z} Z_{W}({\bm t},\bar{{\bm t}},{\bm\xi})=\langle 0|e^{\sum_{k> 0}t_{k}\widehat
{H}_{k}}e^{\sum_{k\in{\mathbb Z}}\xi_{k}\widehat{n}_{k}}e^{\sum_{k\leq
0}\bar{t}_{-k}\widehat{H}_{k}}|W\rangle
\end{eqnarray}
The approach is based on mapping the partition function of a
classical inference problem on a graph onto a calculation of an
expectation value represented by Eq.~(\ref{define-Z}). We have
established such a mapping for some simple Grassmannian models on
planar graphs \cite{08TCC}, where all the details on the suggested
approach will be presented. Note that in the case ${\bm\xi}=0$ and
$\bar{{\bm t}}=0$ the expectation value $Z_{W}({\bm
t},0,0)=\tau_{W}({\bm t})$ is related to the $\tau$-function  of the KP 
integrable hierarchy.

\section{Future Challenges}
\label{sec:Con}

We conclude with a brief and incomplete discussion of future
challenges and opportunities raised by this study.

\begin{itemize}
\item We plan to extend the study looking at new
approximate schemes for intractable planar problems. One new
direction,  suggested in Section \ref{sec:Easy}, consists of
exploring the vicinity of the computationally tractable models reducible
via the BP-gauge transformation to the series of single-connected
loops. It is also of  great interest to explore the vicinity of
integrable tractable models mentioned in
\ref{subsec:comments-fermions}.

\item Perturbative exploration of a larger set of intractable
non-planar problems which are close,  in some sense, to planar
problems, constitutes another interesting extension of the research.
Here,  one would aim to blend the aforementioned planar techniques
with planar (or similar) decomposition techniques,  e.g. these of
the type discussed in \cite{06GJ}.

\item One important component of our analysis consisted in the
Pfaffian re-summation of the single-connected loop (dimer)
contributions, which is a special feature of the graph planarity.
On the other hand, it is known that the planarity is equivalent to
the graph being minor-excluded with respect to  $K_5$ and $K_{3,3}$ subgraphs.
Therefore, one wonders if
there exists a generalization of the Pfaffian reduction to
partition functions of models from other and/or broader graph-minor classes
defined within the graph-minor theory \cite{05Lov}?Likewise, comparing with previous studies of the non-planar/non-spherical cases, based on the dimer approach \cite{n1, n2, n3}.

\item Extending the Loop Series analysis of the binary planar problem
to the q-ary case seems feasible via the Loop Tower construction of
\cite{07CC}. This research should be of a special interest in the
context of recently proposed polynomial quantum algorithm for
calculating partition function of the Potts model \cite{07AAEL}.
Besides, recent progress \cite{DiFrancesco, Kenyon} shows that a
Kasteleyn-type approach is extendable to a $q$-ary case, leading to
the concept of ``heaps of dimers", and (in the continuum limit) to
fascinating connections with special, highly symmetric complex
surfaces, known as Harnack curves.

\item One would also be interested to study  how (and if) phase transitions in the disorder-averaged
planar ensembles, e.g. analyzed in \cite{Sherrington, 83DD, Picco,
Hastings, Tsvelik, Efetov, E-A, FH, SKP}, are related to
distribution of parameters characterizing computational tractability
(complexity) of the models.

\item In \cite{Pascal}, the problem of finding all pseudo-codewords in
a finite cycle code (corresponding to the type of graphical model
discussed in this paper), was addressed by constructing a generating
function known as graph zeta function \cite{Stark}. The interesting
fact discovered in \cite{Pascal} is that this generating function of
pseudo-codewords has a determinant formulation, based on a discrete
graph operator. Hence, one may anticipate an existence of yet
uncovered relation between the graph zeta function and a Pfaffian-Loop
resummation of related graphical models.

\end{itemize}

\section{Acknowledgments}

Research of M.C. and R.T. was carried out under the auspices of the
National Nuclear Security Administration of the U.S. Department of
Energy at Los Alamos National Laboratory under Contract No. DE
C52-06NA25396,  and specifically the LDRD Directed Research grant on
{\it Physics of Algorithms}. M.C. also acknowledges support of the
Weston Visiting Professorship program at the Weizmann Institute of Science,
where he started to work on the manuscript. V.Y.C. acknowledges
support through the start-up funds from Wayne State University.

\begin{appendix}

\section{Graphical Transformations}
\label{sec:Graph}

In this Appendix we discuss graphical transformations reducing any
binary problem to the vertex-function model described by Eq.~(\ref{P_sigma}),
where all vertices are of degree three. Our main focus here is on the planar graphs,  and on
the graphical transformations preserving planarity. However some of
the transformations and considerations discussed below apply to an
arbitrary graph.

\begin{figure}[b]
\begin{center}
    \includegraphics[width=6cm]{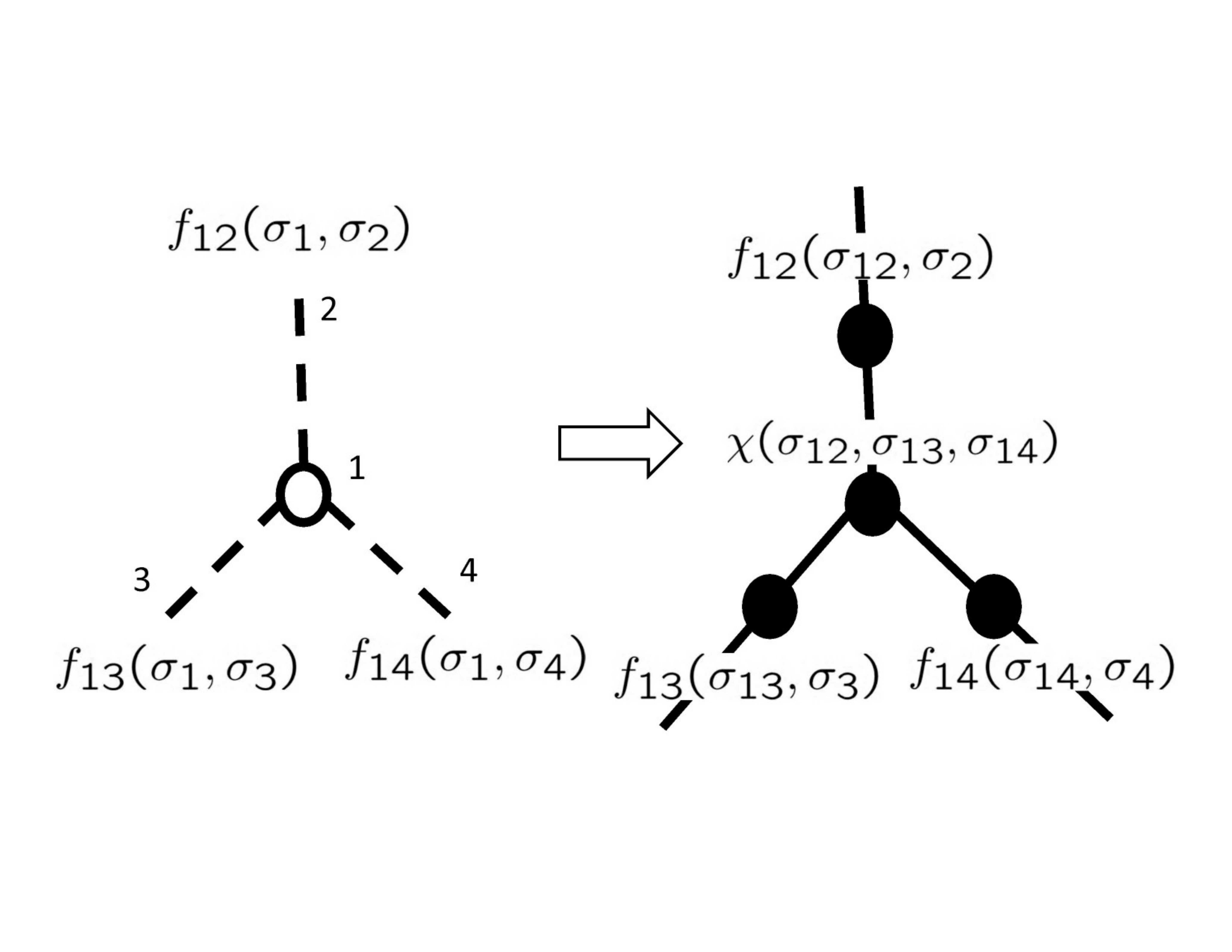}
    \caption{Transformation from binary variable on a vertex, $\sigma_1$,
    to set of variables, $\sigma_{12},\sigma_{13},\sigma_{14}$ on respective
    edges. }
    \label{fig:VtoE}
     \end{center}
\end{figure}

Often the original binary model is not represented in the
vertex-function form. Some or all binary variables describing a
problem may actually be assigned to vertices of a graph, then
respective functions are associated with edges and not vertices.
Obviously, one can also reformulate the model reducing it to the
vertex (canonical for our purposes) form. The transformation is
illustrated in Figure~\ref{fig:VtoE}. Algebraic form of the
transformation shown in the Figure reads,
    $\sum_{\sigma_1}f_{12}(\sigma_1,\sigma_2)f_{13}(\sigma_1,\sigma_3)f_{14}(\sigma_1,\sigma_4)=
     \sum_{\sigma_{12},\sigma_{13},\sigma_{14}}\chi(\sigma_{12},\sigma_{13},\sigma_{14})
     f_{12}(\sigma_{12},\sigma_2)f_{13}(\sigma_{13},\sigma_3)f_{14}(\sigma_{14},\sigma_4)$,
where $\chi(\sigma_{12},\sigma_{13},\sigma_{14})$ is the
characteristic function equal to unity if all variables
$\sigma_{12},\sigma_{13},\sigma_{14}$ are equal each other and equal
to zero otherwise.

\begin{figure}
\begin{center}
    \includegraphics[width=6cm]{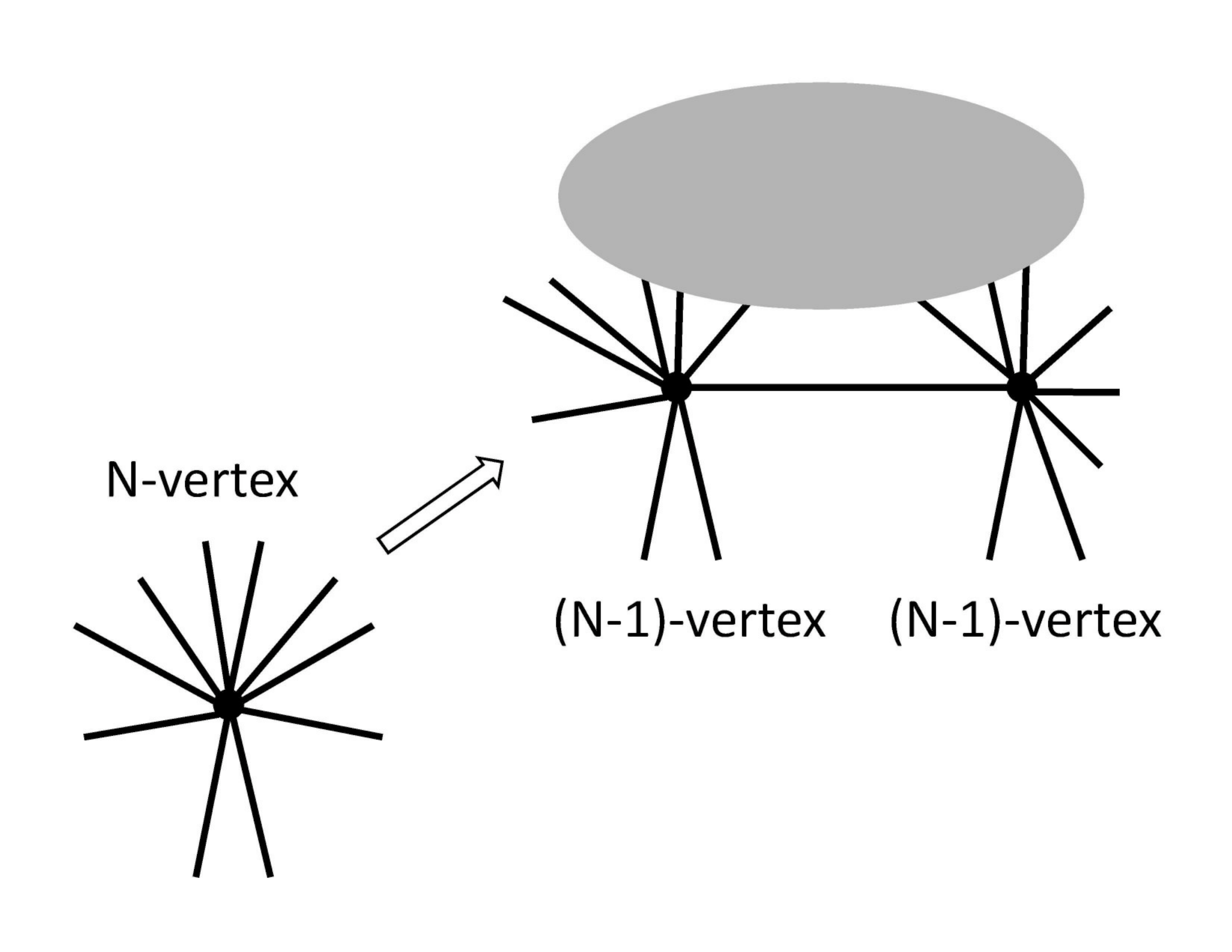}
    \caption{Transformation which allows reduction of an $N$-vertex to two $(N-1)$-vertices.
    It is assumed that (1) number of nodes in the gray area is not large, i.e. $O(N)$,
    (2) the new graph (on the right) is planar, (3) ordering (say clockwise) of the external nodes is
    preserved. The number of parameters characterizing the $N$-vertex is $2^N$ or smaller,
    thus the number of parameters characterizing the two $(N-1)$ vertices and vertices
    from the gray area is sufficient, i.e. $>2\cdot2^{N-1}$, to parameterize the original vertex.}
    \label{fig:NN-1}
     \end{center}
\end{figure}

Next, let us notice that, given a vertex-function model
(\ref{P_sigma}) with the degree of connectivity higher than three,
one can always perform a sequence of transformations reducing the
degree of connectivity of all the nodes in the resulting graphical
model to three. An elementary graphical transformation of the kind
is illustrated in Figure~\ref{fig:NN-1}. It is assumed that the
transformation is applied sequentially to vertices of degree larger
than three till none of these are left. The end result is that: (a)
there are no vertices of degree larger than three left within the
graph; (b) the increase in the total number of vertices is
polynomial; (c) if the original graph is planar the resulting graph
is also planar.

The set of transformations just described is general, and thus often
inefficient, in the sense that knowing specific form of the factor
functions one can practically always do a more efficient, customized and
simpler reduction. Below we will illustrate this point on examples.

\subsection{Ising Model}
\label{subsec:Ising11}

\begin{figure}
\begin{center}
    \includegraphics[width=6cm]{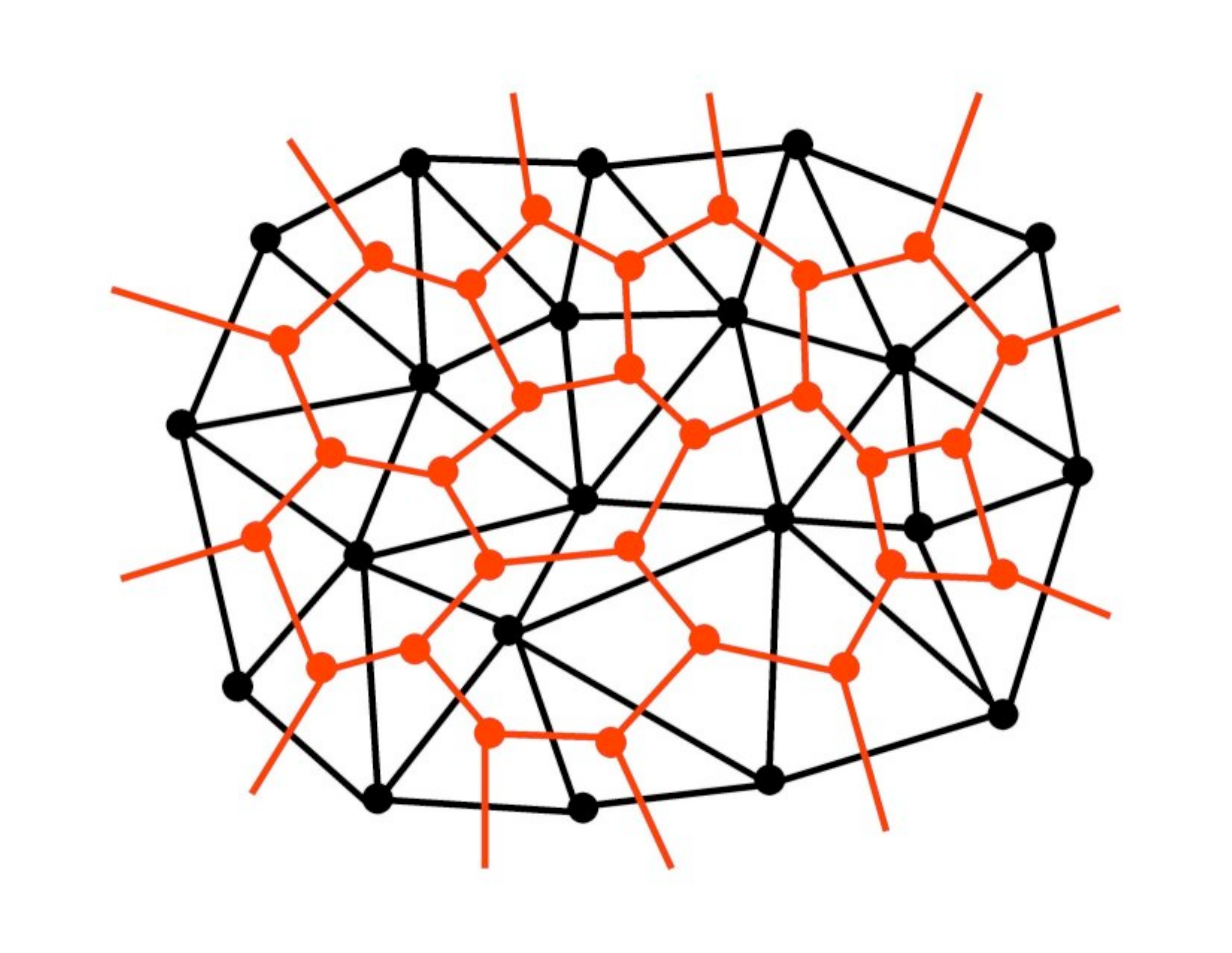}
    \caption{Planar triangulated graph (black) and its dual (red).}
    \label{fig:Triang}
     \end{center}
\end{figure}

The spin glass Ising model is usually defined in terms of
$\sigma_i=\pm 1$ variables associated with vertices of the graph
\begin{eqnarray}
 p({\bm \sigma})=Z^{-1}\exp\left(\sum_{(i,j)}J_{ij}\sigma_i\sigma_j\right),
\label{Ising}
\end{eqnarray}
where summation under the exponential on the r.h.s. goes over all edges
of the graph,  and $J_{ij}$ associated with an edge can be positive
or negative. Obviously one can apply the vertex-to-edges
transformation, explained in Figure~\ref{fig:VtoE}, to restate the
spin glass Ising model as a vertex-function model.   However, in
this case one can also do a simpler transformation to the dual graph.
Let us consider a planar triangulated graph
$\Gamma$ shown in black in Figure~\ref{fig:Triang}.  All vertices of
the respective dual graph, $\Gamma_d$, shown in red in
Figure~\ref{fig:Triang}, have degree of connectivity three. We
assume that the spin glass Ising model is defined on the planar
triangulated graph $\Gamma$. Defining a new variable $\sigma_{ab}$
on an edge of $\Gamma_d$ as the product of two variables of the
original graph $\sigma_{ab}\equiv\sigma_i\sigma_j$ connected by an
edge $(i,j)$ of $\Gamma$ crossing the edge $(a,b$ of $\Gamma_d$, one
finds that the sum on the r.h.s. of Eq.~(\ref{Ising}), rewritten in
terms of the new variables, becomes,
$\sum_{(a,b)\in\Gamma_d}J_{ab}\sigma_{ab}$.  However, the new
variables,  $\sigma_{ab}$ are not independent, but rather related to
each other via a set of local constraints, $\forall a\in\Gamma_d$:
$\prod_b^{(a,b)\in\Gamma_d}\sigma_{ab}=1$. Then, Eq.~(\ref{Ising})
restated in terms of the new variables on the dual graph gets the
following compact vertex-style form
\begin{eqnarray}
 p({\bm \sigma}_d)=Z^{-1}\exp\left(\sum_{(a,b)\in\Gamma_d}J_{ab}\sigma_{ab}\right)
 \prod_{a\in\Gamma}\delta\left(\prod_b^{(a,b)\in\Gamma_d}\sigma_{ab},1\right).
\label{Isingg}
\end{eqnarray}
One interesting observation is that the allowed configurations of
${\bm\sigma}_d\equiv\left(\sigma_{ab}|(a,b)\in\Gamma_d\right)$ on
the dual graph correspond exactly to the single-connected loops on
$\Gamma_d$,  where the loops are built from the excited,
$\sigma_{ab}=-1$, edges. Therefore,  and in accordance with
discussion of Section \ref{sec:SingleLoops}, calculation of the
partition function for the spin glass Ising is reduced to evaluation
of the respective Pfaffian, which is the task of a polynomial complexity.
Notice also that adding a magnetic field (linear in $\sigma$) term
in the expression under the exponent on the r.h.s. of Eq.~(\ref{Isingg})
will raise the complexity level to exponential.

\subsection{Parity-Check Based Error-Correction}
\label{subsec:Parity}

\begin{figure}
\begin{center}
    \includegraphics[width=6cm]{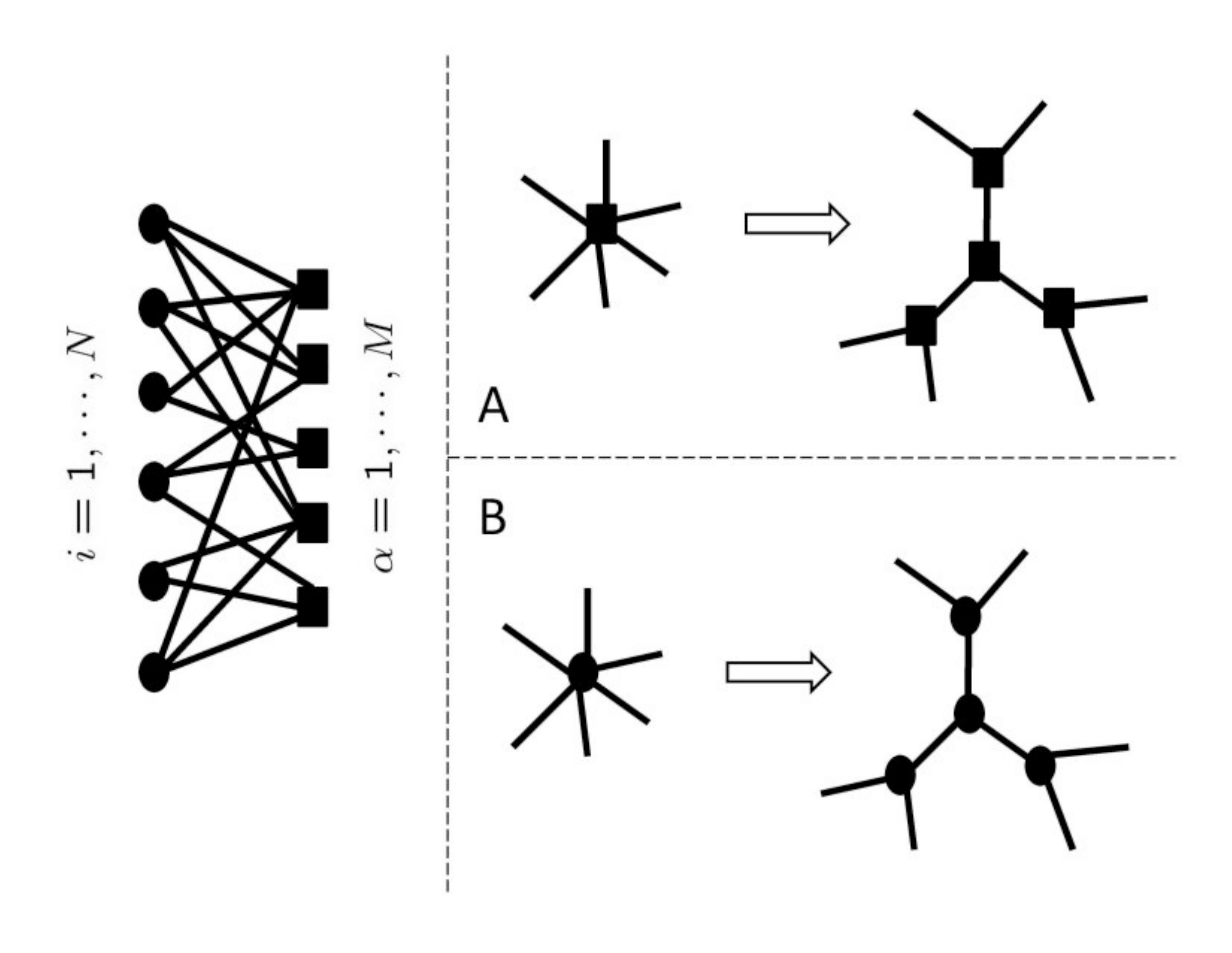}
    \caption{An illustrative example of a Tanner graph (left), as well as check-vertex (A) and bit-vertex
    transformations. }
    \label{fig:LDPC}
     \end{center}
\end{figure}

Consider a linear code with the code-book defined in terms of the
bi-partite Tanner graph, $G=(V_b,V_c,{\cal E})$ consisting of
$N=|V_b|$ bits and $M=|V_c|$ parity checks, and the set of edges
${\cal E}$ relating bits to checks and checks to bits. Then a
message $\vec{\sigma}=(\sigma_i=0,1|i=1,\cdots N)$ is a codeword of
the code if it satisfies all the parity checks, i.e. $\forall
\alpha=1,\cdots M:\ \prod_i^{(i,\alpha)\in{\cal E}}\sigma_i=+1$.
Assuming that all the codewords are equally probable originally, and
that the white channel transform a bit $\sigma$ of the original
codeword into the signal $x$ with the probability $p(x|\sigma)$, one
finds that the probability for $\vec{\sigma}$ to be a codeword
resulted in the measurement $\vec{x}$ is
\begin{eqnarray}
 p(\vec{\sigma}|\vec{x})=\frac{1}{Z}e^{\sum_{i\in V_b}\sigma_i
 h_i}\prod_{\alpha\in V_c}\delta\left(\prod_i^{(i,\alpha)\in{\cal
 E}}\sigma_i,+1\right),\,\,\, h_i\equiv\frac{1}{2}\ln\frac{p(x_i|+1)}{p(x_i|-1)},
 \label{LDPC}
\end{eqnarray}
where, as usual,  the partition function $Z$ is fixed by the
normalization condition,
$\sum_{\vec{\sigma}}p(\vec{\sigma}|\vec{x})=1$.

Eq.~(\ref{LDPC}) represents an example of a mixed graphical model,
with variables $\sigma_i$ defined on bit-vertices,  the parity-check
functions defined on check-vertices and the channel functions
(carrying the dependencies on the log-likelihoods $h_i$) also
associated with the bit-vertices. In this case transformation to the
vertex-style model is done by direct application of the
vertex-to-edges procedure of Figure~\ref{fig:VtoE} to all the
bit-vertices of $G$. Then,  the vertex-style version of
Eq.~(\ref{LDPC}) becomes
\begin{eqnarray}
 && p(\vec{\sigma}|\vec{x})=Z^{-1}\prod_{\alpha\in V_c}
 f_\alpha(\vec{\sigma})\prod_{i\in
 V_b}f_i(\vec{\sigma}_i),\label{LDPC1}\\
 && \forall i:\ \ \vec{\sigma}_i\equiv (\sigma_{i\alpha}=\pm 1|(i,\alpha)\in{\cal
 E}),\\
 &&f_i(\vec{\sigma}_i)=
 \left\{\begin{array}{cc}
 \exp(h_i\sigma_{i\alpha}), & \forall \alpha,\beta \mbox{ s.t. } (i,\alpha),(i,\beta)\in{\cal E}:\ \
 \sigma_{i\alpha}=\sigma_{i\beta},\\
 0, & \mbox{otherwise}\end{array}\right.,\label{fi}\\
 &&
 \forall\alpha:\ \
 \vec{\sigma}_\alpha\equiv(\sigma_{i\alpha}|(i,\alpha)\in{\cal E}),\quad
 f_\alpha(\vec{\sigma}_\alpha)=\delta\left(\prod_i^{(i,\alpha)\in{\cal
 E}}\sigma_{i\alpha},+1\right).
 \label{f_alpha}
\end{eqnarray}

In general, degree of connectivity of bit-vertices and
check-vertices may be arbitrary.  Direct application of the general
procedure explained above (see Figure~\ref{fig:VtoE} and discussion
therein) allows to reduce all the higher-degree nodes to a larger
set of nodes of degree three.  However, a simpler dendro-reduction
is possible both for the bit-vertices and check-vertices. The dendro
trick (e.g. discussed in \cite{07CS} for complexity reduction of a
Linear Programming decoding of LDPC codes) is schematically
illustrated in the two right panels of Figure~\ref{fig:LDPC},  where
respective algebraic relations are
\begin{eqnarray}
 && \mbox{(A)}:\quad \delta\left(\prod_{i=1}^6\sigma_i,+1\right)= \label{LDPC-A} \\
 &&
 \sum_{\sigma_{12},\sigma_{34},\sigma_{56}=\pm 1}\delta(\sigma_1\sigma_2\sigma_{12},+1)
 \delta(\sigma_3\sigma_4\sigma_{34},+1)\delta(\sigma_5\sigma_6\sigma_{56},+1)
 \delta(\sigma_{12}\sigma_{34}\sigma_{56},+1), \nonumber \\
 && \mbox{(B)}:\quad \delta\left(\sigma_1,\cdots,\sigma_6\right)= \label{LDPC-B} \\
 &&
 \sum_{\sigma_{12},\sigma_{34},\sigma_{56}=\pm 1}\delta(\sigma_1,\sigma_2,\sigma_{12})
 \delta(\sigma_3,\sigma_4,\sigma_{34})\delta(\sigma_5,\sigma_6,\sigma_{56})
 \delta(\sigma_{12},\sigma_{34},\sigma_{56}), \nonumber
\end{eqnarray}
and $\delta(\sigma_1,\cdots,\sigma_6)$ is equal to unity if all
arguments are the same, and it is zero otherwise.

\end{appendix}

\end{document}